\documentclass[12pt,twoside,english,prd,nofootinbib,preprint,showpacs,preprintnumbers,floatfix]{revtex4}
\usepackage{mathptmx}
\usepackage{helvet}
\usepackage{color}
\usepackage{courier}
\usepackage[T1]{fontenc}
\usepackage[latin9]{inputenc}
\usepackage[a4paper]{geometry}
\usepackage{url}
\usepackage{babel}

\usepackage{array}
\usepackage{float}
\usepackage{amsmath}
\usepackage{amssymb}
\usepackage{graphicx}

\makeatletter

\providecommand{\tabularnewline}{\\}

\@ifundefined{textcolor}{}
{%
 \definecolor{BLACK}{gray}{0}
 \definecolor{WHITE}{gray}{1}
 \definecolor{RED}{rgb}{1,0,0}
 \definecolor{GREEN}{rgb}{0,1,0}
 \definecolor{BLUE}{rgb}{0,0,1}
 \definecolor{CYAN}{cmyk}{1,0,0,0}
 \definecolor{MAGENTA}{cmyk}{0,1,0,0}
 \definecolor{YELLOW}{cmyk}{0,0,1,0}
}

\makeatother

\begin{document}
\begin{flushright}
MZ-TH/13-046\\
TTK-13-18  
\par\end{flushright}

\vspace{4mm}

\begin{center}
\textbf{\LARGE Higgs CP properties }
\par\end{center}{\LARGE \par}

\begin{center}
\textbf{\LARGE using the tau decay modes at the ILC}{\LARGE{} }
\par\end{center}{\LARGE \par}

\begin{center}
\vspace{6mm}

\par\end{center}

\begin{center}
\textbf{\large Stefan Berge}{\large $^{*}$}%
\footnote{\texttt{\small berge@uni-mainz.de}%
}\textbf{\large {}, Werner Bernreuther}{\large $^{\dagger}$}%
\footnote{\texttt{\small breuther@physik.rwth-aachen.de}%
}\textbf{\large {} and Hubert Spiesberger}{\large $^{*}$}%
\footnote{\texttt{\small spiesber@uni-mainz.de}%
}{\large{} }
\par\end{center}{\large \par}

\begin{center}
$^{*}$ PRISMA Cluster of Excellence, Institut f\"ur Physik (WA THEP),
\\
Johannes Gutenberg-Universit\"at, 55099 Mainz, Germany 
\par\end{center}

\begin{center}
\vspace{-15mm}

\par\end{center}

\begin{center}
$^{\dagger}$ Institut f\"ur Theoretische Physik, RWTH Aachen University,
52056 Aachen, Germany 
\par\end{center}

\begin{center}
\vspace{17mm}
\textbf{Abstract} 
\par\end{center}

We investigate the prospects of determining  the $CP$ nature of the
$126$~GeV neutral spin-0 (Higgs) boson $h$, discovered at the LHC, 
at a future linear $e^{+}e^{-}$ collider (ILC). We consider the 
production of $h$ by the Higgsstrahlung process $e^{+}e^{-}\to Z+ h$ 
and its subsequent decays to $\tau$ leptons, $h \to\tau^{-}\tau^{+}$.
We investigate how precisely a possible pseudoscalar component of 
$h$ can be detected by  the measurement of a suitably defined 
angular distribution, if all major decay modes of the $\tau$ lepton 
are used. From our numerical simulations, we estimate the expected 
precision to the scalar-pseudoscalar mixing angle $\phi$,
 including estimates of the 
background and of measurement uncertainties, to be 
$\Delta\phi \simeq 2.8^{\circ}$ for 
Higgs-boson production at a center-of-mass energy of $250$~GeV and
for a collider with integrated luminosity of $1\,{\rm ab}^{-1}$. 
 
\vspace{35mm}


\newpage{}

\section{Introduction }

The recent results
\cite{Chatrchyan:2012jja,Chatrchyan:2013lba,Aad:2013xqa,Aad:2013wqa}  
by CMS and ATLAS on the production, decays, and properties of the 
neutral boson $h$ of mass $\simeq 126$ GeV that was discovered last
year by these experiments \cite{Aad:2012tfa,Chatrchyan:2012ufa} at 
the LHC support the hypothesis that $h$ is the long-sought Standard
Model (SM) Higgs boson. Nevertheless, much more detailed investigations
will be necessary to firmly establish this expectation. In particular, 
the spin-parity analyses made in $h\to ZZ^*\to 4 l$ 
\cite{Chatrchyan:2012jja,Aad:2013xqa} do not yet prove that $h$ is a 
pure scalar -- they imply that $h$ cannot be a pure pseudoscalar, but 
do not rule out the possibility that it is a mixture of a scalar and 
a pseudoscalar state. It is expected that the profile of this resonance 
can be explored to a large extent at the LHC. 

A high-energy linear $e^+e^-$ collider would be an ideal machine to 
investigate the properties of this spin-0 resonance in great detail, 
i.e., its  decay modes, couplings, and $CP$ parity (and, of course, 
also the properties of other, not too heavy resonances of similar 
type if they exist). For assessing the prospects of exploring this 
particle at a future linear collider, one may revert to the many 
existing phenomenological investigations, within the SM and many of 
its extensions, of Higgs-boson production and decay in $e^+e^-$ 
collisions. As to the prospects of exploring the spin and $CP$ 
properties of a Higgs boson, there have been a number of proposals 
and studies, including
\cite{Dell'Aquila:1985vc,Dell'Aquila:1988rx,Bernreuther:1993df,Soni:1993jc,Barger:1993wt,Hagiwara:1993sw,Kramer:1993jn,Arens:1994nc,Skjold:1994qn,Arens:1994wd,Skjold:1995jp,BarShalom:1995jb,Grzadkowski:1995rx,Gunion:1996vv,Bernreuther:1997af,BarShalom:1997sx,Grzadkowski:1999ye,Choi:2002jk,Bower:2002zx,Desch:2003mw,Desch:2003rw,Rouge:2005iy,Ellis:2005ika,Accomando:2006ga,Bhupal Dev:2007is,Berge:2008wi,Berge:2008dr, Reinhard:2009,Berge:2011ij,Berge:2012wm,Godbole:2011hw,Ellis:2012xd,Ananthanarayan:2013cia,Harnik:2013aja},
that are relevant for  Higgs-boson production and decay at a linear 
collider. 

In this paper,  we apply a method \cite{Berge:2008dr,Berge:2011ij} 
for the determination of the $CP$ nature of a neutral spin-zero 
(Higgs) boson   in its $\tau^+\tau^-$ decays to the production of
$h$ at a future $e^+e^-$ linear collider (ILC). For definiteness, we
consider $e^+e^-\to Z h$, but the analysis outlined below is
applicable to any other $h$ production mode, where 
the $h$ production vertex can be determined. In our analysis, all
major 1-prong and 3-prong $\tau$ decays are taken into account. 
We demonstrate  that the $CP$ nature of $h$ can be determined by this 
approach in a precise and unambiguous way.

\section{Higgs-boson production and decay \label{cross_section} }

The CMS and ATLAS results \cite{Chatrchyan:2013lba,Aad:2013wqa} on 
$h$ production and its couplings to the weak gauge bosons are consistent 
with expectations for the Standard Model Higgs boson. Therefore, the 
$e^{+}e^{-}$  production of $h$ by the Higgsstrahlung process 
\begin{equation}
e^{+}e^{-}\to Z + h  
\label{eeZH_production}
\end{equation}
has a cross section $\sim \sigma_{SM}(Zh)$. We are interested here 
in the decay mode $h\to \tau^-\tau^+$, with subsequent decays 
\begin{equation}
h \to\tau^{-}\tau^{+}\to a^{-}a'^{+}+X\,,\label{phitaudec}
\end{equation}
where $a^{\pm}, a'^{\pm} \in \{e^{\pm}, \mu^{\pm}, \pi^{\pm}, 
a_{1}^{L,T,\pm}\}$ and $X$ denotes neutrinos and $\pi^0$.

The interaction of a Higgs boson $h$ of arbitrary $CP$ nature to 
$\tau$ leptons is described by the Yukawa Lagrangian 
\begin{equation}
{\cal L}_{Y} = 
-(\sqrt{2}G_{F})^{1/2} m_{\tau}
\left(a_{\tau}\bar{\tau}\tau+b_{\tau}\bar{\tau}i\gamma_{5}\tau\right) h\,,
\label{YukLa}
\end{equation}
where $G_{F}$ denotes the Fermi constant and $a_{\tau}$, $b_{\tau}$
are the reduced dimensionless $\tau$ Yukawa coupling constants. In 
order to be able to compare with other studies in the literature, we 
use in the following  sections, instead of  (\ref{YukLa}), the equivalent 
parameterization
\begin{equation}
{\cal L}_{Y} = 
- g_\tau
\left(\cos\phi\bar{\tau}\tau + \sin\phi\bar{\tau}i\gamma_{5}\tau\right)
h \,,
\label{YukLa-phi}
\end{equation}
where $g_\tau$ is the effective strength of the Yukawa interaction and
$\phi$ describes the degree of mixing of the scalar and pseudoscalar 
component: 
\begin{equation}
\label{Yuparphi}
g_\tau 
=
(\sqrt{2}G_{F})^{1/2}m_{\tau}\sqrt{a_{\tau}^{2}+b_{\tau}^{2}}
\, , 
\qquad
\tan\phi=\frac{b_{\tau}}{a_{\tau}}
\, .
\end{equation}
In the following sections, we take into account the main 1- and 3-charged prong 
$\tau$ decay modes:
\begin{eqnarray}
\tau & \to & l+\nu_{l}+\nu_{\tau}\,, 
\qquad l=e,\mu \, ,  \label{taulept} 
\\
\tau & \to & \pi+\nu_{\tau}\,, \label{taupi} 
\\
\tau & \to & \rho+\nu_{\tau}\to\pi+\pi^{0}+\nu_{\tau}\,,
\label{taurho} 
\\
\tau & \to & a_{1}+\nu_{\tau}\to\pi+2\pi^{0}+\nu_{\tau}\,, 
\label{taua1} 
\\
\tau & \to & a_{1}^{L,T}+\nu_{\tau}\to2\pi^{\pm}+\pi^{\mp}+\nu_{\tau}\,.
\label{taua1LT}
\end{eqnarray}
The decay mode (\ref{taua1LT}), in fact a 3-prong $\tau$ decay with 
three charged pions, will also be called `1-prong'
because the track, i.e., the 4-momentum of the $a_1^\pm$ resonance 
can be obtained from the 4-momenta of the three charged pions. 
Moreover, by using known kinematic distributions, the longitudinal 
$(L)$ and transverse $(T)$ helicity states of the $a_1$ resonance 
can be separated \cite{Rouge:1990kv,Davi93,Kue95,Stahl:2000aq}.
Thus, the $\tau^-\tau^+$ decays that we analyze are 
of the form (\ref{phitaudec}) with $a, a'$ as specified below
(\ref{phitaudec}).

The observables that we use to determine the $CP$ nature of $h$ in its
$\tau$ decays are based on $\tau$-spin correlations
\cite{Bernreuther:1993df,Bernreuther:1997af,Berge:2008wi,Berge:2008dr,Berge:2011ij}. 
The charged lepton $l=e,\mu$ in (\ref{taulept}), the charged pion
in (\ref{taupi}) - (\ref{taua1}), and the  $a_{1}^{L,T}$ in 
(\ref{taua1LT}) act as $\tau$-spin analyzers. The 
$\tau$-spin analyzing power is maximal for the direct decays to pions, 
$\tau^\mp \to \pi^\mp$, and for $\tau^\mp \to  a_{1}^{L,T,\mp}$. For 
$a_{1}^{L-}$ and $a_{1}^{T-}$ it is $+1$ and $-1$,
respectively. 
For the decays (\ref{taulept}),
 (\ref{taurho}), and  (\ref{taua1}), the $\tau$-spin analyzing power 
 of  $l^\mp$ and $\pi^\mp$     depends on the
energy of these  particles,
cf. Appendix~A. We will apply 
cuts  on the respective  energy
to optimize the $\tau$-spin analyzing power.

The differential cross section of the  
production process~(\ref{eeZH_production})
and subsequent decay~(\ref{phitaudec}) can be derived from  
Eq.~(4) of~\cite{Berge:2011ij}. For a Higgs boson of arbitrary 
 $CP$ nature it is given by
\begin{eqnarray}
\label{eq:dsigma_1}
 d\hat{\sigma} & = & N_{ \tau} 
\overline{\left|M_{a^{-}a'^+}\right|}^{2} d\Omega_{Z} \, d\Omega_{\tau}\,\,
dE_{a^{-}}d\Omega_{a^{-}}
dE_{a'^{+}}d\Omega_{a'^{+}} /(2\pi) 
\\[1ex]
 &\times& 
 n\left(E_{a^-}\right)  n\left(E_{a'^+}\right)  \{
  A  - b\left(E_{a^-}\right) \!b\left(E_{a'^+}\right) \!
   \left[ c_1 {\bf\hat{q}}^{-} \cdot{\bf\hat{q}}^{+} +c_2 {\bf\hat{k}}\cdot
       {\bf\hat{q}}^{-} {\bf\hat{k}} \cdot {\bf\hat{q}}^{+} 
          +c_3 {\bf\hat{k}} \cdot ({\bf\hat{q}}^{-}\times
          {\bf\hat{q}}^{+}) \right]\! \} ,
\nonumber 
\end{eqnarray}
where 
\[ N_{ \tau} = \frac{\sqrt{2} G_F m_{\tau}^2
  \beta_{\tau}}{128\pi^{3}s}, \qquad  
\overline{\left|M_{a^{-}a'^+}\right|}^{2}=\overline{\sum}\left|M\left(e^-e^+ \to Zh \right)\right|^{2}
\left|D^{-1}\left(h\right)\right|^{2} \mbox{B}_{_{\tau^{-}\to
    a^{-}}} \mbox{B}_{_{\tau^{+}\to a'^{+}}} \, . \]
 In Eq.~(\ref{eq:dsigma_1}),
 ${\bf\hat{k}}$ denotes
  the normalized $\tau^-$ momentum in the Higgs-boson 
 rest frame and  ${\bf\hat{q}}^-$  $({\bf\hat{q}}^+)$ 
 is  the $a^-$  $(a'^+)$ direction of flight in the 
 $\tau^-$  $(\tau^+)$ rest frame. The functions $n$ and $b$ are
  defined in Appendix~A,
 the coefficients $A,\,c_i$ are given in Table~I of~\cite{Berge:2011ij},
 $\beta_{\tau}$ is the $\tau$ velocity, $s=E_{cm}^2$, and $D^{-1}$
 is the Higgs-boson propagator. 

Choosing  ${\bf\hat{k}}$ to be the $z$ axis of a right-handed coordinate 
  system
 and integrating 
 Eq.~(\ref{eq:dsigma_1})  over the polar angles
$d\theta_{a'^{-}}$ and $d\theta_{a^{+}}$, we obtain:
\begin{eqnarray}
\label{eq:dsigma_2}
 d\hat{\sigma} & = & N_{ \tau}  
\overline{\left|M_{a^{-}a'^+}\right|}^{\,2} d\Omega_{Z} \, d\Omega_{\tau}\,\,
dE_{a^{-}} \, dE_{a'^{+}} 
 d\varphi  \Big[ v+ u \cdot  \cos \left( \varphi - 2 \phi \right)
 \Big] \, ,
\end{eqnarray}
where 
\[ \varphi = \phi_{a^{-}}-\phi_{a'^{+}} \, , \qquad 0\le \varphi\le
2\pi \, , \]
is the difference of the azimuthal angles of $a^{-}$ and $a'^{+}$,
\[ u =- n\left(E_{a^-}\right)
b\left(E_{a^-}\right) n\left(E_{a'^+}\right) b\left(E_{a'^+}\right)
 \frac{\pi^2 p^2_h}{8} \frac{g_{\tau}^2}{\sqrt{2}G_F m_{\tau}^2}
 \, , \qquad v = 4  n\left(E_{a^-}\right) n\left(E_{a'^+}\right)\,A \,
 , \]
and $\phi$ is the Higgs mixing angle defined in (\ref{Yuparphi}).

The distribution (\ref{eq:dsigma_2}) holds also in the $\tau\tau$ zero-momentum frame
(ZMF). The angle $\varphi$ is equal to the angle between the {\it
  signed} normal vectors of the $\tau^-\to a^-$ and $\tau^+\to a'^+$
decay planes spanned by the unit vectors ${\bf\hat k}$, ${\bf\hat{q}}^{-}$ and  
$-{\bf\hat k}$, ${\bf\hat{q}}^{+}$, respectively.
 Instead of  determining $\varphi$ in the $\tau\tau$ ZMF one can 
 measure this angle also in the zero-momentum frame of the
 charged prongs $a^{-}$ and $a'^{+}$ (cf.~\cite{Berge:2008dr} and
 below). This has the advantage that the  $\tau^\mp$ momenta need not
 be reconstructed.

\section{Method and Observables \label{Observables} }

Our method to determine the $CP$ nature of $h$ requires, in the case of 
the 1-prong $\tau^-\tau^+$ decays (\ref{phitaudec}), the measurement 
of the 4-momenta of the charged prongs $a^-$, $a'^+$ and their impact 
parameter vectors (unit vectors) ${\bf \hat{n}}_{\mp}$ in the 
laboratory frame. The 4-vectors $n_{\mp}^{\mu} = (0,{\bf \hat{n}}_{\mp})$ 
are then boosted into the $a^- a'^+$ zero-momentum frame
(ZMF). The 
spatial parts of the resulting 4-vectors  $n_{\mp}^{*\mu}$ are 
decomposed into their normalized components 
${\bf\hat{n}}_{\textbar\textbar}^{*\mp}$ and ${\bf\hat{n}}_{\perp}^{*\mp}$ 
that are parallel and perpendicular to the respective $a^-$ and $a'^+$ 
3-momentum. 
With  this prescription, one determines in the $a^- a'^+$ ZMF
   the {\it unsigned} normal vectors
${\bf\hat{n}}_{\perp}^{*-}$ and  
 ${\bf\hat{n}}_{\perp}^{*+}$ of the $\tau^-\to a^-$ and 
 $\tau^+ \to a'^+$   decay planes. The distribution of 
the angle between these
two planes~\cite{Berge:2008dr},
\begin{equation}
\varphi^{*} = \arccos({\bf \hat{n}}_{\perp}^{*+} \cdot
{\bf \hat{n}}_{\perp}^{*-}) \, ,
\label{phistar}
\end{equation}
where $0\le  \varphi^{*} \le \pi$, 
discriminates between  $CP=\pm 1$  Higgs boson states. 
The simultaneous 
measurement of (\ref{phistar}) and of the  $CP$-odd and $T$-odd triple 
correlation
\begin{equation} 
\label{CP-oddTrip}
{\cal O}_{CP}^{*} = 
{\bf \hat{q}}_{-}^{*}\cdot({\bf \hat{n}}_{\perp}^{*+} \times 
{\bf \hat{n}}_{\perp}^{*-}) \, ,
\end{equation}
where  ${\bf \hat{q}}_{-}^{*}$ is the normalized $a^{-}$ momentum in 
the $a^{-} a'^{+}$ ZMF, allows for an unambiguous determination of the
$CP$ nature of $h$ \cite{Berge:2008dr}. If $h$ is a mixture of a 
$CP$-even and -odd state, the distribution of (\ref{CP-oddTrip}) is 
asymmetric with respect to ${\cal O}_{CP}^{*} = 0$. 
In order to determine the ratio $b_\tau/a_\tau$ of the reduced Yukawa 
couplings (\ref{YukLa}) or, equivalently, the mixing angle $\phi$ 
defined in (\ref{Yuparphi}), one would fit theoretical predictions 
for $\sigma^{-1}d\sigma/d\varphi^{*}$
and
$\sigma^{-1}d\sigma/d{\cal O}_{CP}^{*}$ to the corresponding measured
distributions. In addition, associated asymmetries can be measured. 
Some results of this approach, applied to the reactions 
(\ref{eeZH_production}), (\ref{phitaudec}), were presented in the 
workshop report \cite{Berge:2012wm}. \\

Here we use a slight variation 
of our approach just outlined. Instead of using both the distribution 
of the `unsigned' angle $\varphi^{*}$, Eq.\ (\ref{phistar}), which is defined in the range 
$0\leq \varphi^{*}\leq \pi$, and of ${\cal O}_{CP}^{*}$, 
the same information is of course contained  in the distribution of the `signed' angle 
between  the $\tau^-\to a^-$ and 
 $\tau^+ \to a'^+$   decay planes in the $a^- a'^+$ ZMF.
This angle  which will be called $\varphi_{CP}^{*}$ in the following and that varies between $0$ and 
$2\pi$ is obtained by the following prescription:
\begin{equation}
\varphi_{CP}^{*} = 
\left\{ \begin{array}{ccc}
\varphi^{*} & if & {\cal O}_{CP}^{*}\geq 0 \, ,
\\
2\pi-\varphi^{*} & if & {\cal O}_{CP}^{*}<0 \, .
\end{array} \right.
\label{phistar_CP}
\end{equation}
In terms of this angle the  triple correlation (\ref{CP-oddTrip}) ${\cal
  O}_{CP}^{*} = \sin \varphi^{*}_{CP}$. The distribution of
(\ref{phistar_CP})
 is given by (\ref{eq:dsigma_2}) with $\varphi \to  \varphi^{*}_{CP}.$

\begin{figure}[t]
\includegraphics[height=6.2cm]{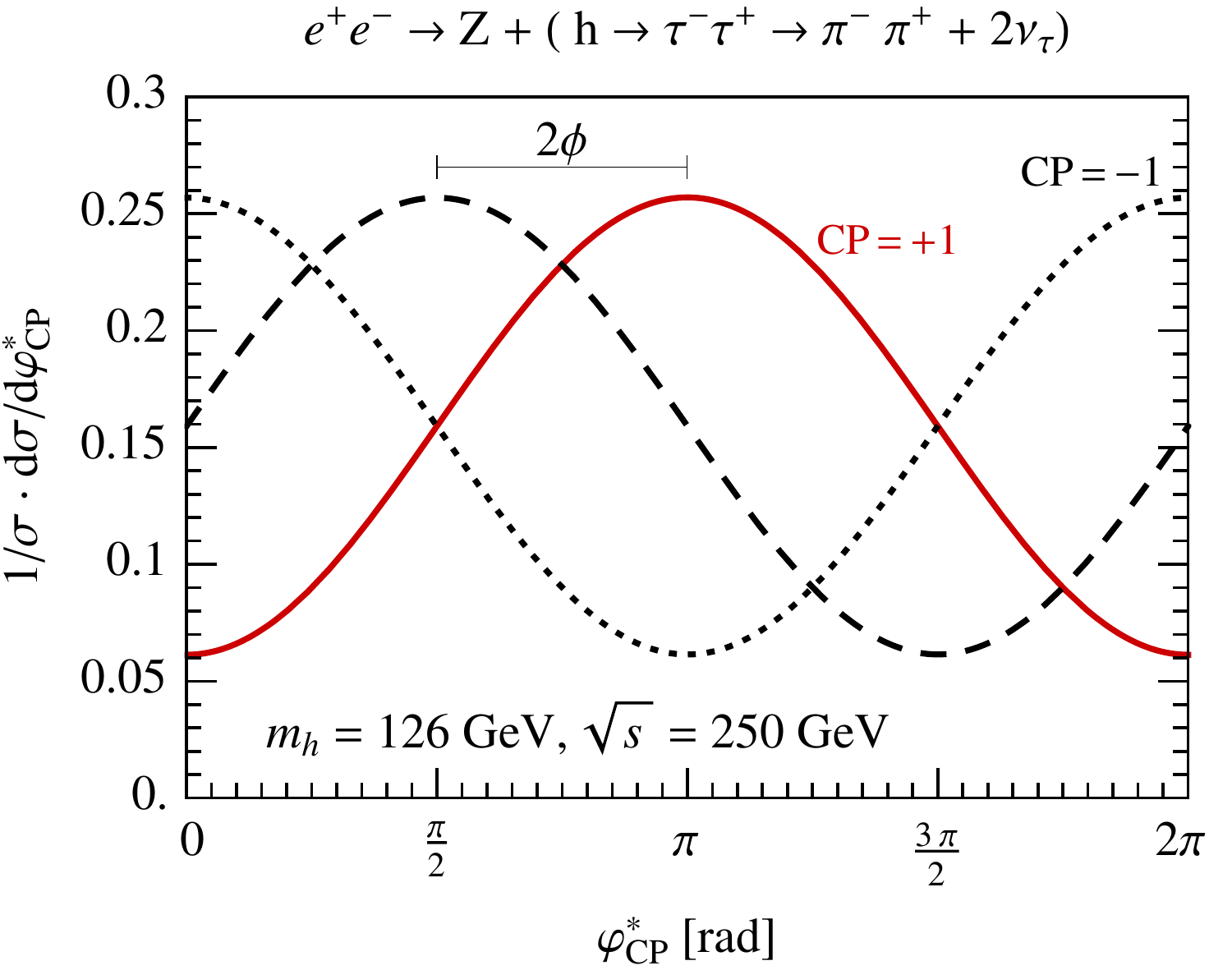}\hspace*{.0cm}
\includegraphics[height=6.2cm]{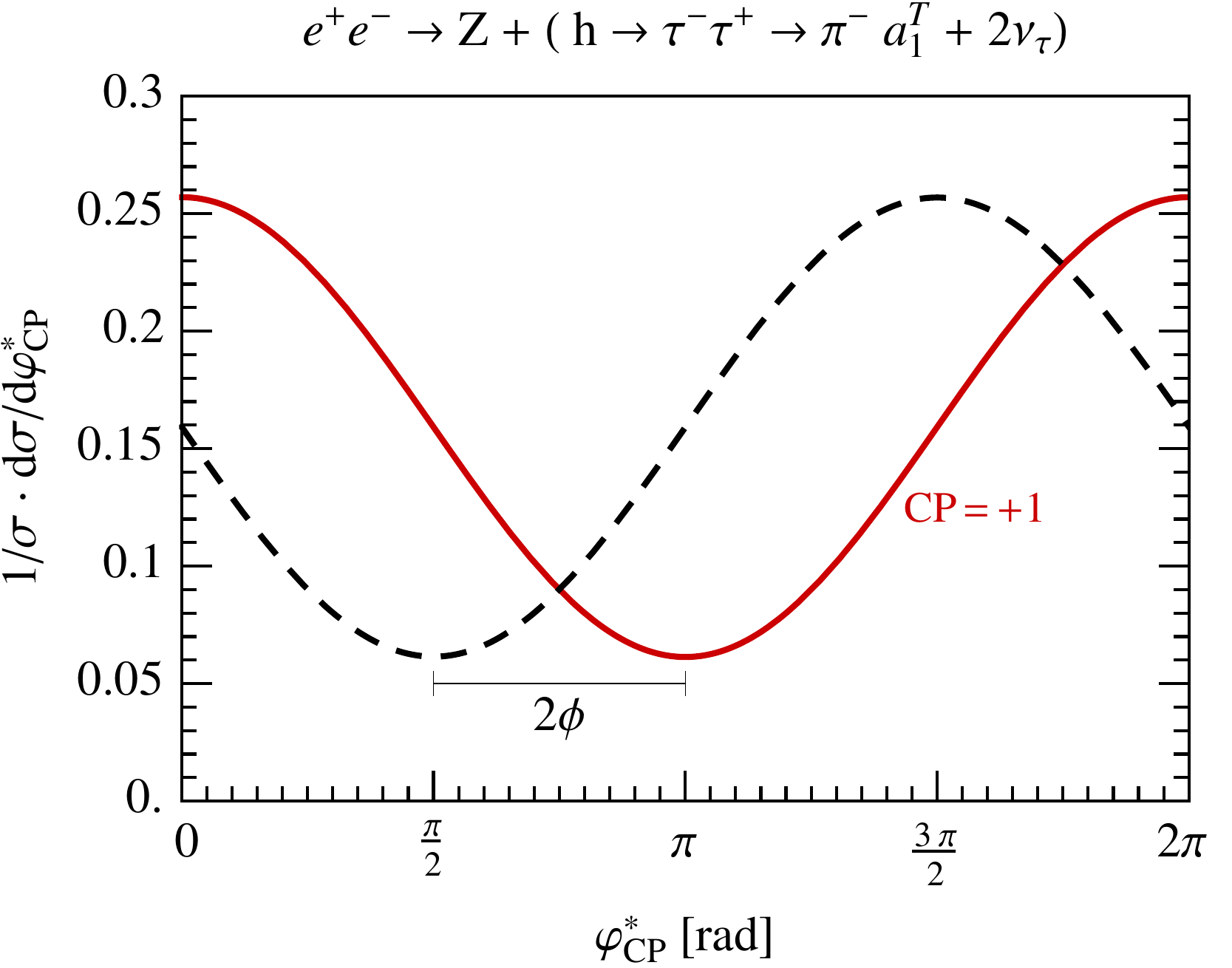}
\caption{Left side:
  Normalized distribution of $\varphi_{CP}^{*}$ for $\tau^-\tau^+ \to 
  \pi^{+}\pi^{-}+2\nu$. The red solid line, the black dotted line, and 
  the black dashed line show the distributions for a  $CP$-even
  Higgs boson ($\phi=0$), for a  $CP$-odd boson ($\phi=\pm{\pi}/{2}$), 
  and for a $CP$ mixture with $\phi=-\frac{\pi}{4}$, respectively. 
 Right side: The normalized distribution of $\varphi_{CP}^{*}$ 
      for $\tau^-\tau^+ \to \pi^-a_1^{+T} +2\nu$ for a $CP$-even
  Higgs boson and  a $CP$ mixture with $\phi=-\frac{\pi}{4}$.
}
\label{fig:h_pipi_detcuts}
\end{figure}

In order to illustrate the discriminating power of (\ref{phistar_CP}), 
we consider the $h\to \tau^-\tau^+\to\pi^-\pi^+$ decay mode. The 
normalized distribution of $\varphi_{CP}^{*}$ for this decay channel 
is shown on the left side of  Fig.~\ref{fig:h_pipi_detcuts}.
The red solid line shows the distribution for a pure $CP$-even Higgs
boson ($\phi=0$) and the  black dashed line results from the decay of
an ideal $CP$ mixture (with $\phi=-{\pi}/{4}$). The distance between 
the maxima of the distributions for the $CP$-even Higgs boson and the 
$CP$ mixture is given by  $2\phi$. For completeness, the 
distribution resulting from the decay of a  $CP$-odd state ($\phi=\pm{\pi}/{2}$) is also shown.
 
For each decay channel $aa'$ as specified above, the Higgs mixing angle 
$\phi$ can be obtained from the measured differential distributions 
by fitting the function $f=u \cos\left(\varphi_{CP}^{*}-2\phi\right)+v$
(subject to the constraint $\int_0^{2\pi}d\varphi_{CP}^{*} f\, = \, 
2\pi v\,=\,\sigma_{aa'}$) to the measured distribution of 
$\varphi_{CP}^{*}$. For a fixed number of $aa'$ events the sensitivity 
to the mixing angle $\phi$ of this channel depends on the product of
the $\tau$-spin analyzing powers of $a$ and $a'$. We find it convenient 
to define  the following asymmetry that embodies this fact:
\begin{equation}
A^{aa'} 
= \frac{1}{\sigma_{aa'}} 
[
\sigma_{aa'}(u \cos(\varphi_{CP}^{*}-2\phi)>0) 
- 
\sigma_{aa'}(u \cos(\varphi_{CP}^{*}-2\phi)<0) ] 
\,=\,\frac{-4u}{2\pi v} \, .
\label{phiCP_asym}
\end{equation}
The value of $A^{aa'}$ does not depend on the mixing angle $\phi$, 
but it reflects the combined strengths of the $\tau$-spin analyzing 
powers of $a$ and $a'$. Thus, the larger $A^{aa'}$, the smaller the 
statistical error $\Delta\phi$ for a given number of events. The 
asymmetry is largest if both $\tau$ leptons decay either directly to 
$\pi^\pm$ or to three charged prongs, i.e., to $a_1^{\lambda}$ 
$(\lambda=L,T)$. For these decays $|A^{aa'}|=39.3\%$.

In the next section, we determine the statistical uncertainty 
$\Delta\phi$ of the mixing angle in the following way 
\cite{Desch:2003rw}. We assume an integrated collider 
luminosity of 1\,ab$^{-1}$. For some value of $\phi$, for instance
$\phi=-\pi/8$, we generate for each decay channel the differential 
distribution of $\varphi_{CP}^{*}$, using $20$ bins between $0$ and 
$2\pi$. Then we perform a fit to this distribution with the 3-parameter 
function $u \cos\left(\varphi_{CP}^{*}-2\phi\right)+v$. This yields
a certain value of $\phi$. We repeat this procedure a $1000$ times, 
then  fit a Gaussian curve to the resulting  distribution of $\phi$ 
and take the width of the Gaussian  as the expected statistical  
uncertainty $\Delta\phi$.

As already mentioned above, the $\tau$-spin analyzing power is maximal 
for the direct decays $\tau^\mp\to \pi^\mp$ and for $\tau^\mp \to 
a_{1}^{L,T \mp}.$ The  $\tau$-spin analyzing power of the charged 
lepton in $\tau^\mp\to l^\mp$ and of the charged pion from $\tau^\mp 
\to\rho^\mp$ and $\tau^\mp\to a_1^\mp$ can be enhanced by applying an 
appropriate cut on the energy of the lepton and the pion, respectively 
(see Figs.\ 1a and 4 in \cite{Berge:2011ij}). Ideally, this cut should 
be applied in the $\tau^\mp$ rest frames. The reconstruction of these 
rest frames is, however, not possible for the leptonic $\tau$ decay 
modes.

For $\tau^-\tau^+ 
\to a^-a'^+ + \nu_\tau{\bar\nu}_\tau$, $a, a'= \rho, a_1$, the rest 
frames of $\tau^-$ and $\tau^+$ can be reconstructed by solving 
kinematic constraints \cite{Tsai:1965hq}, a technique that is well known 
in $\tau$ physics at $e^+e^-$ colliders (cf., e.g., \cite{Stahl:2000aq} 
for a review). Therefore, we apply for the decays $\tau\tau \to \rho
\rho,a_1a_1, \rho a_1$ an appropriate cut on the energy of the charged 
pion from  $\rho$ and $a_1$ decay in the respective $\tau$ rest frame. 
Because the $\tau$-spin analyzing functions 
$b(E_{\pi}^{\tau-rest})$ are slowly varying functions of the energy 
$E_{\pi}^{\tau-rest}$, uncertainties in these reconstructed energies 
due to measurement uncertainties will not have a dramatic effect on 
our conclusions below.
 
This method does not work if one $\tau$, or both, decay to leptons.
In this case, we use the fact that the 4-momentum of the $Z$ boson in 
the reaction (\ref{eeZH_production}) can be determined unambiguously 
from its visible decay modes. In this case the rest frame of the Higgs 
boson $h$ is also known. For the decays $\tau^-\tau^+ \to l^-l'^+, 
l^\mp \pi^\pm$, where the pion results from  $\rho$ or $a_1$ decay, we 
apply cuts on the energies of the charged lepton and pion in the rest 
frame of $h$, in order to enhance their $\tau$-spin analyzing powers 
and in order to analyze whether or not such cuts increase the sensitivity 
of these decay modes to the mixing angle $\phi$ in a significant way.

At this point we briefly recall several other detailed studies on the 
determination of the  $CP$ parity of a Higgs boson in its $\tau^+ \tau^-$ 
decays at a linear collider. In Refs.\ 
\cite{Bower:2002zx,Desch:2003mw,Desch:2003rw} the hadronic 1-prong decay 
$\tau \to \rho \nu$ was analyzed. The observable used there, namely the
acoplanarity angle of the $\rho^+$ and $\rho^-$ decay planes, requires 
the reconstruction of the $\rho^+ \rho^-$ ZMF, i.e., the measurement of 
the $\pi^\pm$ and the $\pi^0$ momenta. In addition, the reconstruction 
of approximate $\tau^\pm$ rest frames is needed, because the method of 
\cite{Desch:2003rw} requires event selection that involves the knowledge 
of the $\pi^\pm$ and the $\pi^0$ energies in these frames. The authors 
of Ref.\ \cite{Desch:2003rw} conclude that at a linear collider with 
integrated luminosity of 1\,ab$^{-1}$ the scalar-pseudoscalar mixing 
angle can be measured with an uncertainty $\Delta\phi=6^\circ$. 
Background $\tau\tau$ studies were not made in this analysis.

In Ref.~\cite{Reinhard:2009} the hadronic $\tau$-decay channels
$\tau\to\pi,\rho,a_{1}$ were analyzed. (A similar analysis was made
before by \cite{Rouge:2005iy} for $\tau^\mp \to\pi^\mp \nu$.)  The observable used in this 
study is the difference of the azimuthal angles of the polarimeter 
vectors of the decays $\tau^\mp \to \pi^\mp,\rho^\mp, a_1^\mp$.
The measurement of this observable requires the reconstruction of 
the $\tau$ rest frames. The study includes a full detector simulation, 
takes into account the relevant background from $ZZ$ production and
determines appropriate kinematical cuts for suppressing this background.
The  resulting signal-to-background ratio is  estimated to be
$S/B=4.5$.  Assuming an integrated luminosity of 300\,{\rm fb}$^{-1}$, the authors 
of \cite{Reinhard:2009} conclude that a mixing angle of 
$\phi=-\pi/8$  can be excluded with $4.5\sigma$ with respect to the 
$\phi=0$ hypothesis.

The recent proposal \cite{Harnik:2013aja} for determining the mixing
angle $\phi$, both at the LHC and ILC, is based on the decay mode
 $\tau^\mp \to\rho^\mp \nu$ and relies on a complete reconstruction
 of the $\rho^\mp$ 4-momenta.

\section{Numerical Results}

For our analysis we use a Higgs boson mass of $m_{h}=126$~GeV. We 
assume  that the strength of the $ZZh$ vertex is as predicted by the 
SM. If $h$ has a pseudoscalar component, this component would not 
couple to  $ZZ$ at tree level. As expected from multi-Higgs extensions 
of the SM, it is likely that in this case the coupling of the scalar 
component of $h$ to $ZZ$ is reduced with respect to the SM coupling. 
A coupling of the pseudoscalar 
component of $h$ to $ZZ$ can be induced at the loop-level but, using 
the results of \cite{Bernreuther:2010uw} and current 
LHC results, one concludes that such a coupling 
must be very small as compared to the respective coupling of the scalar 
component. The decay mode $h\to\tau\tau$ is, however, already  affected 
at tree level if $h$ is a $CP$ mixture. 

We assume in the following that the 
$126$~GeV Higgs boson is SM-like, i.e., is predominantly a
scalar.
 Therefore we use  both 
 for the cross section  of the Higgsstrahlung process
 (\ref{eeZH_production})
 and the branching ratio $B(h\to\tau\tau)$ the
 respective SM predictions (c.f., for instance, \cite{Dittmaier:2011ti}).
Furthermore, we assume that 
the  electron and positron beams  have a longitudinal polarization of 
$-0.8$ and $0.3$, respectively. Then we obtain $\sigma = 
325$\,fb at $\sqrt{s} = 250$~GeV for the cross section of (\ref
{eeZH_production}) and use this value to estimate the number of 
events in the various decay channels.

Our method \cite{Berge:2008dr}
 requires that the Higgs production vertex is known.
 It can be determined from the charged tracks of the $Z$-boson decay
products. In the case of
  leptonic $\tau$ decay modes the application of energy cuts
 to enhance the $\tau$-spin analyzing power requires also the
  knowledge of the  Higgs-boson 4-momentum (see below).
   Therefore, only the visible $Z$ boson decay modes are taken 
into account. In fact, we only  consider 
\begin{equation} 
\label{Ztolight}
Z \to e^- e^+, ~ 
\mu^- \mu^+, ~ 
u{\bar u}, ~ 
d{\bar d}, ~ 
s{\bar s} \, .
\end{equation}
The other decays ($Z\to\tau^- \tau^+$, $c{\bar c}$, $b{\bar b}$) would 
require a more detailed study.
For instance, $Zh$ production
   with $Z\to b{\bar b}, h\to {\tau\tau}$ has to be disentangled from
  $Zh,Z\to {\tau\tau}, h \to b{\bar b}$.
  The branching ratio of the 
decays (\ref{Ztolight})   is $B(Z \to {\rm light})=0.5$~\cite{Beringer:1900zz}.

We analyze the reactions (\ref{eeZH_production}), (\ref{phitaudec})
in the narrow width approximation both for the Higgs boson and the 
$\tau$ leptons.  In this section 
we do not apply cuts for the selection of the signal 
events. An estimate of the effect of  such cuts will be done in Sec.~\ref{sec:finest}.
The branching fractions of the $\tau^-\tau^+$ decays that we 
use in the following sections are summarized in 
Table~\ref{tab:tautau_Branching-fractions} \cite{Beringer:1900zz}.

The cross sections for the various signal reactions that we use, e.g.,
$Zh$ with $Z\to{\rm light}$ and $h\to\tau\tau\to l l'$, are denoted by
$\sigma_0^{ll}$, etc., in the Tables below. For $\tau \to l, \rho,
a_1$, we apply energy cuts in order to enhance the $\tau$-spin
analyzing power. 
 The resulting reduced cross section
 ratios $\sigma/\sigma_0$ are also given in 
Tables~\ref{tab:A_sigma_ll} -- \ref{tab:A_sigma_rhoa1} below.
 
For each decay channel $aa'$, the value of the asymmetry 
(\ref{phiCP_asym}) is also given in these Tables. Note, however, 
that we use this asymmetry only as a qualitative measure for the 
sensitivity to the scalar-pseudoscalar mixing angle $\phi$. The 
estimates of the statistical uncertainty $\Delta\phi$ were made for 
the various  channels as described above (see text after 
Eq.~(\ref{phiCP_asym})). The estimates given in this
 section  are based on 
an integrated luminosity of 1\,ab$^{-1}$.

\begin{table}
\begin{tabular}{|c|>{\centering}p{1.7cm}|>{\centering}p{1.7cm}|>{\centering}p{2.7cm}|>{\centering}p{2.7cm}|>{\centering}p{2.7cm}|}
\hline 
Branching ratios $[\%]$ & $l^{-}$ & $\pi^{-}$ & $\rho^{-}\to\pi^{-}\!+\!\pi^{0}$ & $a_{1}^{-}\to\pi^{-}\!+\!2\pi^{0}$ & $a_{1}^{L,T}\to2\pi^{-}+\pi^{+}$\tabularnewline
\hline 
$l^{+}$ & $12.4$ & $7.7$ & $18.0$ & $6.6$ & $6.3$\tabularnewline
\hline 
$\pi^{+}$ &  & $1.2$ & $5.6$ & $2.0$ & $2.0$\tabularnewline
\hline 
$\rho^{+}\to\pi^{+}+\pi^{0}$ &  &  & $6.5$ & $4.7$ & $4.6$\tabularnewline
\hline 
$a_{1}^{+}\to\pi^{+}+2\pi^{0}$ &  &  &  & $0.9$ & $1.7$\tabularnewline
\hline 
$a_{1}^{L,T+}\to\pi^{-}+2\pi^{+}$ &  &  &  &  & $0.8$\tabularnewline
\hline 
\end{tabular}
\caption{
  Branching fractions of the different decay modes \cite{Beringer:1900zz}. 
  The off-diagonal elements are the sum of the branching fractions of the 
  respective decay mode and the charge-conjugated mode, for instance,
  $B_{l^{-}\pi^{+}}+B_{l^{+}\pi^{-}}$.
}
\label{tab:tautau_Branching-fractions}
\end{table}
%

\subsection{$\tau^-\tau^+\to
  \{\pi^-,a_1^{-\lambda}\}\{\pi^+,a_1^{+\lambda'}\}$ }

First we consider the case where both $\tau$ leptons decay either
directly to a charged pion or to $a_1$ with a subsequent 3-prong decay,
and we assume that the polarization state of the $a_1$, i.e.,
$a_1^{\pm,\lambda}$ ($\lambda  =L,T$) can be reconstructed efficiently
at a linear collider \cite{Rouge:1990kv,Davi93,Kue95,Stahl:2000aq}. 
As a measure of the expected efficiency one may consider the sensitivity
 of the decay distributions of $\tau\to \pi, \rho, a_1$ to the $\tau$
polarization. If one uses all the information about the respective
hadronic final state, the sensitivity computed from the theoretical
distributions  is the same for the three decay modes ($S=0.58$) 
  \cite{Davi93}.
For $\tau$ pair production  at LEP and subsequent semi-hadronic decays of both
  $\tau$ leptons, it was shown by the ALEPH collaboration
\cite{Heister:2001uh} that experimentally $S\simeq
0.48$ can be achieved, which amounts to a degradation of only about $17\%$. It seems
 not too optimistic to assume that such an efficiency can also be
achieved -- and probably improved -- at a future $e^+e^-$ collider.
Therefore we assume that in the case of both $\tau$ leptons decaying
semi-hadronically, the spin-analyzing power  of $a_1$ is maximal and the
 asymmetry  
(\ref{phiCP_asym}) takes its maximal value $|A^{aa'}|=39.3\%$ for all 
combinations of $a,a'=\pi,a_1^\lambda$. 
 The resulting distributions of 
$\varphi_{CP}^{*}$ for $aa'=\pi\pi$ are shown in 
Fig.~\ref{fig:h_pipi_detcuts} (left part). Identical distributions are obtained for 
$aa' = a_1^{\lambda} a_1^{\lambda}$, $\pi a_1^L$. For $aa' = 
\pi a_1^T, a_1^L a_1^T$ the corresponding distributions are shifted by 
$\varphi_{CP}^{*}\to \varphi_{CP}^{*}+\pi$ (see 
Fig.~\ref{fig:h_pipi_detcuts}, right part), because the 
$\tau^\mp$-spin analyzing power of $a_1^{\mp T}$ is $\mp 1$, i.e., 
opposite to the analyzing powers of $\pi^\pm$ and $a_1^{\mp L}$.

 To minimize $\Delta\phi$ it is therefore essential 
to disentangle the polarization states  
$a_1^\lambda$ in an efficient way. The cross sections of the $\pi\pi$, 
$a_1^{\lambda}a_1^{\lambda'}$, and $\pi a_1^{\lambda}$ channels are 
$\sigma_0^{\pi\pi}=0.12$~fb, $\sigma_0^{a_1^{\lambda}a_1^{\lambda'}} 
=0.08$~fb, and $\sigma_0^{\pi a_1^{\lambda}}=0.20$~fb, respectively. 
Adding the three channels yields  $400$ events for a luminosity 
of 1~ab$^{-1}$. Using the procedure described above we estimate the 
statistical error on the mixing angle from fitting the 
$\varphi_{CP}^{*}$ distribution of these decay modes to be $\Delta\phi = 3.3^\circ$.

\subsection{ $\tau^-\tau^+\to l^- l'^+$ }

The $\tau^-\tau^+\to l^- l'^+$ decay mode has a rather large branching 
fraction of $12.4\%$. 
 However, 
the spectral function $b(E_{l})$
  changes sign at $E_{l} = 
m_{\tau}/4$ (see Fig.\ 1b of Ref.\ \cite{Berge:2011ij}).
The inclusive $\tau$-spin 
analyzing power of $l=e,\mu$, that is, the spin-analyzer quality that 
results from integrating over the lepton energy $E_l$ in the $\tau$ rest frame,  
 is therefore small. Thus   the  $\varphi_{CP}^{*}$ 
distribution  of the $l^- l'^+$ channel  is not very 
sensitive to the mixing angle $\phi$ if no energy cuts 
are made.

Because the $\tau$ rest frames cannot be reconstructed for these
channels, we try to enhance the $\tau$-spin analyzing powers by 
applying the cuts $E_{l^-} = E_{l^+} > E_{cut}$ on the lepton
energies in the Higgs-boson rest frame. Our results for this channel 
are given in the left part of Table~\ref{tab:A_sigma_ll}. 
The asymmetry  is
  increased slightly by these cuts  -- however, this 
growth is outweighed by the decrease of the cross section.
We conclude that the statistical error $\Delta\phi$ cannot be reduced 
in a significant way by applying a cut on the energy of the lepton(s) 
in the Higgs-boson rest frame. The sensitivity to $\phi$ of these 
channels is rather poor.

\begin{table}
\begin{tabular}{|>{\centering}p{1.8cm}|>{\centering}p{1.4cm}|>{\centering}p{1.5cm}|>{\centering}p{1.2cm}||>{\centering}p{1.8cm}|>{\centering}p{2.4cm}|>{\centering}p{2.7cm}|>{\centering}p{1.4cm}|}
\hline 
$E_{l^{-}}=E_{l^{+}}$ in Higgs rest frame  & $A^{ll'}$  $[\%]$ & $\sigma/\sigma_{0}$ 

$\sigma_{0}^{ll'}=1.25$ fb & $\Delta\phi$ $[^\circ]$ & $E_{l}$ in Higgs rest frame  & $A^{l\pi}(A^{la_{1}^{L}})$ $[\%]$ 

 & $\sigma/\sigma_{0}$ 

$\sigma_{0}^{l\pi}=0.78$ fb

($\sigma_{0}^{la_{1}^{\lambda}}=0.64$ fb)& $\Delta\phi$  $[^\circ]$
for $l\pi+l a_{1}^{\lambda}$\tabularnewline
\hline 
$>0$ GeV & $4.4$ & $1$ & $20.4$ & $>0$~GeV & $-13.1$ $(-13.1)$ & $1$ &  $5.2$\tabularnewline
\hline 
$>10$ GeV & $6.5$ & $0.54$ & $19.6$ & $>10$~GeV & $-16.0$ ($-16.0)$ & $0.74$ $(0.74)$ & 

$5.0$\tabularnewline
\hline 
$>15$ GeV & $7.6$ & $0.37$ & $19.6$ & $>15$~GeV & $-17.2$ ($-17.2)$ & $0.61$ $(0.61)$ & 

$5.5$\tabularnewline
\hline 
$>20$ GeV & $8.6$ & $0.24$ & $22.1$ & $>20$~GeV & $-18.3$ $(-18.3)$ & $0.5$ $(0.5)$ & 

$5.4$\tabularnewline
\hline 
\end{tabular}
\caption{
  Asymmetries, cross sections, and uncertainties $\Delta\phi$ for the 
  decay modes $\tau\tau\to ll'$ (left part) and $\tau\tau\to l\pi$, 
  $l a_1^{\lambda}$, $(\lambda=L,T)$ (right part) for several cuts 
  on the energy of the lepton(s) in the Higgs-boson rest frame. 
}
\label{tab:A_sigma_ll}
\end{table}
%

\subsection{$\tau^-\tau^+ \to l^\mp\{\pi^\pm,a_1^{\pm\lambda}\}$}

The results given in the right 
part of Table~\ref{tab:A_sigma_ll}  show  whether or not 
 a cut on the energy of the charged lepton 
in the Higgs-boson rest frame leads 
to a higher sensitivity to 
the mixing angle $\phi$ for these  decay modes.  
A  cut of $E_{l^-} = E_{l^+} > 10$~GeV
 reduces the uncertainty  $\Delta \phi$. For larger energy cuts,
the  increase in spin-analyzing power
is completely offset by the decrease of the cross section. 
 In any case,  the sensitivity to $\phi$ of 
these decay 
modes  is higher than   the sensitivity of the 
$\tau^-\tau^+\to l^- l'^+$ decay channels.

\subsection{$\tau^-\tau^+\to l^\mp\{\rho^\pm,a_1^{\pm}\}$}

The inclusive $\tau$-spin analyzing power of the charged pion
from $\rho$ decays and from the 1-prong $a_1$ decays is very small.
As in the case of the lepton this is due to the fact that the
corresponding spectral functions  $b(E_{\pi^\pm})$
change sign
within the physical region, see Fig.~4 of Ref.\ \cite{Berge:2011ij}.
For these decay modes, where the decay of $a_1$ involves only one 
charged pion and  the final state contains three neutrinos, the 
reconstruction of the $\tau^\pm$ rest frames is not possible. 
Therefore, we investigated the effect of cuts on the energies of 
the charged lepton and pion in the Higgs-boson rest frame. The 
resulting sensitivities to the mixing angle are given in 
Table~\ref{tab:A_sigma_lrho_la1}. 
 We observe that for the $l^\pm\rho^\mp$ decay channel,
a cut on the charged pion energy is 
essential to obtain a reasonable sensitivity to $\phi$. E.g., with a minimum 
cut of $E_{\pi^\pm}>25$~GeV an asymmetry of $-10.2\%$ 
 is obtained. Similarly,  the asymmetry of the 
$l^\pm a_1^\mp$ decay channel can be increased by a small amount. 
However, even with such cuts the overall sensitivity to $\phi$
of these decay modes is low. 

\begin{table}
\begin{tabular}{|>{\centering}p{2.3cm}|>{\centering}p{1.4cm}|>{\centering}p{1.8cm}|>{\centering}p{1.4cm}||>{\centering}p{2.3cm}|>{\centering}p{1.5cm}|>{\centering}p{1.8cm}|>{\centering}p{1.4cm}|}
\hline 
$E_{\pi}$ in Higgs rest frame ($E_{l}>10$~GeV)  & $A^{l\rho}\,\,[\%]$  & $\sigma/\sigma_{0}$

$\sigma_{0}^{l\rho}=$ \\ $1.81$ fb  & $\Delta\phi$ $[^\circ]$  & $E_{\pi}$ in Higgs rest frame ($E_{l}>10$~GeV)  & $A^{la_1}\,\,[\%]$  & $\sigma/\sigma_{0}$

$\sigma_{0}^{la_{1}}=$ \\ $0.66$ fb & $\Delta\phi$  $[^\circ]$
\tabularnewline
\hline 
$>0$~GeV  & $0.7$  & $0.74$  &  $-$ & $>0$~GeV  & $2.8$  & $0.74$ & $37.3$\tabularnewline
\hline 
$>25$~GeV  & $-10.2$  & $0.22$  & $15.6$  & $>25$~GeV  & $-3.4$  & $0.16$ & $43.9$\tabularnewline
\hline 
$<25$~GeV  & $5.4$  & $0.53$  & $19.2$  & $<25$~GeV  & $4.4$  & $0.59$ & $32.5$\tabularnewline
\hline 
$<20$~GeV & $6.4$  & $0.46$  & $16.7$  & $<20$~GeV  & $5.0$  & $0.50$ & $31.5$\tabularnewline
\hline 
$<15$~GeV & $7.2$  & $0.375$  & $16.4$  & $<15$~GeV & $5.3$  & $0.39$ & $32.5$\tabularnewline
\hline 
\end{tabular}
\caption{
  Asymmetries, cross sections, and uncertainties $\Delta\phi$ for the 
  decays $\tau\tau\to l\rho$ (left part) and $\tau\tau \to la_{1} \to 
  l+\pi+2\pi^{0}$ (right part). $E_{\pi}$ and $E_{l}$ denote the energy 
  of the charged pion and lepton in the Higgs-boson rest frame. The cut
  $E_{l}\ge 10$~GeV has been applied. The values given for
  $\sigma_{0}^{l\rho}$ and $\sigma_{0}^{la_{1}}$ were computed without
  this cut.
}
\label{tab:A_sigma_lrho_la1}
\end{table}
%

\subsection{$\tau^-\tau^+\to \rho^-\rho^+$, $a_1^-a_1^+$ }
\label{suse:rhorho}

For the double-hadronic decay channels, we assume that the $\tau$ rest
frames can be reconstructed~\cite{Tsai:1965hq}.
The ALEPH experiment at LEP showed \cite{Heister:2001uh} that this is
possible with an efficiency of $80\%$ for this type of decay modes.
 Therefore, a cut on the
energy of the charged pion from the decay of $\rho$ or $a_1$ can be applied
in the respective $\tau$ rest frame.

\begin{table}
\begin{tabular}{|>{\centering}p{2.5cm}|>{\centering}p{1.3cm}|>{\centering}p{1.6cm}|>{\centering}p{1.4cm}||>{\centering}p{3cm}|>{\centering}p{1.4cm}|>{\centering}p{1.6cm}|}
\hline 
$E_{\pi}$ in $\tau$ rest frame in GeV  & $A^{\rho\rho}$ $[\%]$  & $\sigma/\sigma_{0}$

$\sigma_{0}^{\rho\rho}=0.66$ fb  & $\Delta\phi$   $[^\circ]$  & $E_{\pi}$ in $\tau$ rest frame in GeV  & $A^{a_{1}a_{1}}$

 $[\%]$  & $\sigma/\sigma_{0}$

$\sigma_{0}^{a_{1}a_{1}}=0.09$ fb \tabularnewline
\hline 
 & $0.1$  & $1$  &  &  & $1.2$  & $1$\tabularnewline
\hline 
$E_{\pi^{+}},E_{\pi^{-}}>0.6$  & $25.9$  & $0.21$  & $9.2$  & $E_{\pi_{1}^{+}},E_{\pi_{1}^{-}}>0.53$  & $5.1$  & $0.1$\tabularnewline
\hline 
$E_{\pi^{+}},E_{\pi^{-}}<0.6$  & $23.5$  & $0.3$  & $7.9$  & $E_{\pi_{1}^{+}},E_{\pi_{1}^{-}}<0.53$  & $6.8$  & $0.47$\tabularnewline
\hline 
$E_{\pi_{1}}>0.6>E_{\pi_{2}}$  & $-24.6$  & $0.49$  & $5.9$  & $E_{\pi_{1}}>0.53>E_{\pi_{2}}$  & $-5.9$  & $0.43$\tabularnewline
\hline 
combined  & $24.5$  & $1$  & $4.1$  & combined  & $6.2$  & $1$\tabularnewline
\hline 
\end{tabular}
\caption{
  Asymmetry, cross section, and uncertainty $\Delta\phi$ for the decay
  $\tau\tau\to\rho\rho$ (left part). $E_{\pi}$ denotes the energy of 
  the charged pion in the  respective $\tau$ rest frame. The right part 
  of the Table contains the asymmetry and cross section for  $\tau\tau 
  \to a_{1}a_{1}$. The sensitivity of this decay mode to $\phi$ is very 
  low; therefore $\Delta\phi$ is not given.
}
\label{tab:A_sigma_rhorho_a1a1}
\end{table}

The left part of Table~\ref{tab:A_sigma_rhorho_a1a1} contains our
results for $\tau\tau\to\rho\rho$. Without a cut on the energies
of $\pi^\pm$ from the decays of $\rho^\mp$, this decay mode is useless for
the determination of $\phi$, as indicated by the small value
$A^{\rho\rho} = 0.1\%$. Judiciously chosen cuts on $E_{\pi^{-}}$ and
$E_{\pi^{-}}$, however, lead to a dramatic increase in sensitivity 
to $\phi$. As the three sets of cuts given in 
Table~\ref{tab:A_sigma_rhorho_a1a1} do not intersect, one can compute 
their combined effect. We find that the mixing angle can be determined 
with an uncertainty $\Delta\phi\simeq 4^\circ$ in this decay mode.

The right part of Table~\ref{tab:A_sigma_rhorho_a1a1} contains the
asymmetry and cross section for  $\tau\tau\to a_{1}a_{1}$ for 
three sets of cuts on the energies $E_{\pi^{-}}$, $E_{\pi^{+}}$ 
of the charged pions from the 1-prong $a_1$ decays. The increase 
in $\tau$-spin analyzing power by these cuts is offset by the decrease 
of the cross section which is small already without cuts. The 
sensitivity of this decay mode to $\phi$ is very low; therefore 
$\Delta\phi$ is not given.

\subsection{$\tau^-\tau^+\to \rho^\pm a_{1}^{\pm}$}

Table~\ref{tab:A_sigma_rhoa1} contains our results for the 
`non-diagonal' 1-prong hadronic decay mode, $\tau\tau\to\rho a_{1} 
\to \pi^+\pi^-$. Various sets of cuts were applied to the energies 
of the charged pions from the $\rho$ and $a_1$ decay in the respective 
$\tau$ rest frames. As the four sets of cuts given in
Table~\ref{tab:A_sigma_rhoa1} do not intersect, we can compute the
resulting  combined statistical uncertainty $\Delta\phi$, and we 
obtain 
 $\Delta\phi\simeq 10^\circ$.

\begin{table}
\begin{tabular}{|>{\centering}p{4.5cm}|>{\centering}p{2.4cm}|>{\centering}p{3.4cm}|>{\centering}p{3.2cm}|}
\hline 
$E_{\pi}$ in $\tau$ rest frame in GeV  & $A^{\rho a_{1}}$ $[\%]$  & $\sigma/\sigma_{0}$

$\sigma_{0}^{\rho a_{1}}=0.48$ fb   & $\Delta\phi$ $[^\circ]$
\tabularnewline
\hline 
 & $0.38$  & 1  & \tabularnewline
\hline 
$E_{\pi(\rho)}>0.6$, $E_{\pi(a_{1})}>0.53$  & $11.5$  & $0.14$  & $32.0$\tabularnewline
\hline 
$E_{\pi(\rho)}<0.6$, $E_{\pi(a_{1})}<0.53$  & $12.6$  & $0.38$  & $18.6$\tabularnewline
\hline 
$E_{\pi(\rho)}>0.6$, $E_{\pi(a_{1})}<0.53$  & $-13.3$  & $0.31$  & $19.1$\tabularnewline
\hline 
$E_{\pi(\rho)}<0.6$, $E_{\pi(a_{1})}>0.53$  & $-11.0$  & $0.17$  & $30.0$\tabularnewline
\hline 
combined  & $12.4$  & $1$  & $10.4$\tabularnewline
\hline 
\end{tabular}
\caption{
\label{tab:A_sigma_rhoa1}
  Asymmetry, cross section, and uncertainty $\Delta\phi$ for the 
  channel $\tau\tau\to\rho a_{1}$. 
}
\end{table}
%

\subsection{$\tau^-\tau^+ \to\{\rho^\mp, a_{1}^{\mp} \} \{\pi^\pm,
  a_{1}^{\pm\lambda}\}$}

Finally, we consider the double-hadronic decay channels where
one of the $\tau$ leptons decays either directly to a pion or to three 
charged prongs via $a_1^\lambda$ $(\lambda=L,T$). We assume that the 
polarization states $a_1^\lambda$ can be reconstructed also in these
channels. Notice that the sum of the branching ratios of the $\rho\pi$ 
and $\rho a_{1}^{T,L}$ decay modes is 
 almost  twice as large as that 
of the $\rho\rho$ mode analyzed in Sect.~\ref{suse:rhorho}. In addition, 
we recall that the $\tau$-spin analyzing power is maximal for $\tau 
\to \pi/a_{1}^{L,T}$. Therefore we expect that the sensitivity to the 
mixing angle $\phi$ of these two channels, $\rho\pi$ and 
$\rho a_{1}^{T,L}$, will be better than for the $\rho\rho$ channel 
given in Sect.~\ref{suse:rhorho}, if an appropriate cut on the energy 
of the charged pion from the $\rho$ decay in the respective $\tau$ 
rest frame is applied. 

Our results for the $\rho\pi$ and $\rho a_{1}^{T,L}$ decay channels 
are given in the left part of Table~\ref{tab:A_sigma_rho_a1_pi_a1L}.
Because the pion-energy cuts of the third and fourth row of this Table 
do not intersect, we can compute their combined effect. We find that
with these decay channels the mixing angle can be determined with 
a statistical uncertainty $\Delta\phi\simeq 3^\circ$.

The right part of Table~\ref{tab:A_sigma_rho_a1_pi_a1L} contains 
our results for the $a_1\pi$ and $a_1 a_{1}^{T,L}$ decay modes.
Because both the asymmetries $A^{a_1\pi}$, $A^{a_1a_1^\lambda}$ and
the cross sections are smaller than in the case where one $\tau$
decays to a $\rho$, one can achieve only a moderate sensitivity to 
$\phi$ with these decay channels. 

\begin{table}
\begin{tabular}{|>{\centering}p{1.5cm}|>{\centering}p{1.7cm}|>{\centering}p{2.4cm}|>{\centering}p{1.6cm}||>{\centering}p{1.5cm}|>{\centering}p{1.8cm}|>{\centering}p{2.2cm}|>{\centering}p{1.6cm}|}
\hline 
$E_{\pi(\rho)}$ in $\tau$ rest frame in GeV  & $A^{\rho\pi}$ $(A^{\rho a_{1}^{L}})$
  $[\%]$  & $\sigma/\sigma_{0}$

$\sigma_{0}^{\rho\pi}=0.56$fb

$(\sigma_{0}^{\rho a_{1}^{\lambda}}=0.46{\rm fb})$  & $\Delta\phi$ $[^\circ]$

for $\pi+a_{1}^{\lambda}$  & $E_{\pi(a_{1})}$ in $\tau$ rest frame in GeV  & $A^{a_{1}\pi}\,(A^{a_{1}a_{1}^{L}})$

 $[\%]$  & $\sigma/\sigma_{0}$

$\sigma_{0}^{a_{1}\pi}=0.21$fb ($\sigma_{0}^{a_{1}a_{1}^{\lambda}}=0.17$fb)  & $\Delta\phi$ $[^\circ]$

 for $\pi+a_{1}^{\lambda}$\tabularnewline
\hline 
 & $-2.1$ $(-2.1)$  & 1  &  &  & $-6.9$ $(-6.9)$  & $1$  & $24.2$\tabularnewline
\hline 
$>0.6$  & $\,\,\,\,31.9$ $(\,\,\,\,\,\,31.8)$  & $0.45$ $(0.45)$  & $3.8$  & $>0.53$  & $\,\,\,\,\,14.2$ $(\,\,\,\,\,14.2)$  & $0.31$ $(0.31)$  & $19.8$\tabularnewline
\hline 
$<0.6$  & $-30.3$ $(-30.4)$  & $0.55$ $(0.55)$  & $3.7$  & $<0.53$  & $-16.3$ $(-16.4)$  & $0.69$ $(0.69)$  & $11.1$\tabularnewline
\hline 
combined  & $31.0$

$(31.0)$  & $1$  & $2.6$  & combined  & $15.7$

$(15.7)$  & $1$  & $9.1$\tabularnewline
\hline 
$>0.7$  & $\,\,\,\,36.0$ $(\,\,\,\,\,\,35.9)$  & $0.36$ $(0.35)$  & $3.7$  & $<0.4$  & $-22.3$ $(-22.4)$  & $0.41$ $(0.41)$  & $10.1$\tabularnewline
\hline 
$<0.5$  & $-31.9$ $(-31.9)$  & $0.48$ $(0.48)$  & $3.7$  & $>0.6$  & $\,\,\,\,\,20.6$ $(\,\,\,\,\,20.5)$  & $0.19$ $(0.19)$  & $18.5$\tabularnewline
\hline 
\end{tabular}
\caption{
  Left part: Asymmetries, cross sections, and uncertainties 
  $\Delta\phi$ for the decays $\tau\tau\to\rho\pi,$ $\rho
  a_{1}^{\lambda}$ $(\lambda=L,T)$. The energy of the charged pion 
  from a $\rho$ decay in the respective $\tau$ rest frame is denoted 
  by $E_{\pi(\rho)}$. Right part: Asymmetries, cross sections, and 
  uncertainties $\Delta\phi$ for the decays $\tau\tau\to a_1\pi,$ 
  $a_1 a_{1}^{\lambda}$ $(\lambda=L,T)$.
}
\label{tab:A_sigma_rho_a1_pi_a1L}
\end{table}
%

\section{Uncertainties}
\label{sec:uncert}

There are a number of uncertainties that will affect the measurement
of  the distributions $\varphi_{CP}^{*}$ in the various decay channels.
One source of uncertainty results from beamstrahlung and initial state
radiation (ISR) which  change the total cross section. While 
beamstrahlung effects on the Higgsstrahlung process 
(\ref{eeZH_production}) are small (for $m_{h}=126$~GeV and at
$\sqrt{s}=250$~GeV), ISR reduces the cross section by a few
percent~\cite{Barger:1993wt}.

The normalized $\varphi_{CP}^{*}$ distribution is not 
directly affected by beamstrahlung and ISR, because the quality of the 
distribution depends, first of all, on the precision with which the 
normalized impact parameter vectors ${\bf\hat n}_\perp^\pm$ can be 
measured. For the lepton-lepton and lepton-hadron decay channels, we 
found the effect of ISR on the 
 normalized
$\varphi_{CP}^{*}$ distribution to be 
negligibly small, even if cuts on the final lepton or pion energies 
in the reconstructed Higgs-boson rest frame are applied. (The Higgs-boson 
rest frame is reconstructed using $\vec{p}_{h}=-\vec{p}_{Z}$.) On the 
other hand, as discussed above, for some of the double hadronic decay 
modes, it is  necessary to reconstruct the $\tau$ rest frames 
for applying energy cuts that enhance the sensitivity to the $CP$ mixing 
angle $\phi$. Beamstrahlung and ISR will affect the quality of this 
reconstruction and will therefore reduce the sensitivity to $\phi$ to 
some extent~\cite{Reinhard:2009}.

\begin{figure}[H]
\noindent 
\centering{}\includegraphics[scale=0.6]{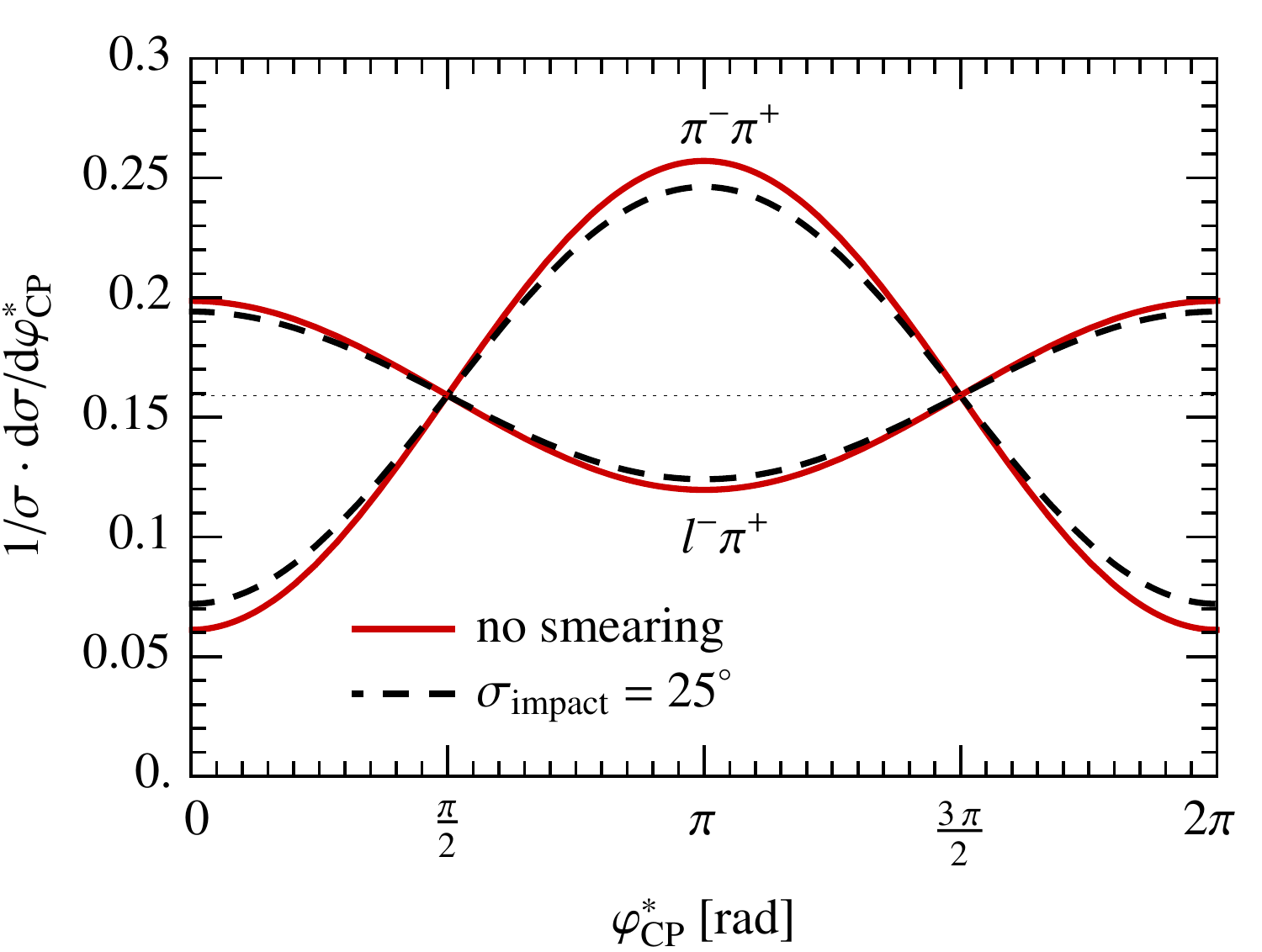}
\caption{
  The normalized $\varphi_{CP}^{*}$ distributions for the decays of a 
  $CP$-even Higgs boson $h \to \tau\tau \to \pi\pi$ and $h \to 
  \tau\tau \to l\pi$ if the impact parameter direction is smeared
  by a Gaussian with $\sigma_{impact}=25^{\circ}$. The solid line
  shows the distributions without smearing.
}
\label{fig:Smearing}
\end{figure}

The largest uncertainty involves the measurement uncertainty
 of  the impact parameter vectors 
${\bf\hat n}_\perp^\pm$. In order to study this issue
we have performed a Monte Carlo simulation taking into account the expected 
measurement uncertainties by smearing the impact parameter direction. 
Measurement errors were assumed to be described by a Gaussian with a 
$1\sigma$ uncertainty of $\sigma_{impact}=25^{\circ}$, as suggested 
in Ref.\ \cite{Desch:2003rw}. The resulting effect on the 
$\varphi_{CP}^{*}$ distributions is shown in Fig.~\ref{fig:Smearing} 
for a $CP=+1$ state and for the decay modes $h \to \tau^-{\tau}^+ \to \pi^{-}
\pi^{+} + \nu_{\tau} + \bar{\nu}_{\tau}$ and $h \to \tau^-{\tau}^+ \to 
l^{-}\pi^{+}+\nu_{\tau}+\bar{\nu}_{\tau}+\bar{\nu}_{l}$. The effect 
turns out to reduce the asymmetry by a factor of $\approx 0.9$.
 We expect that a more 
realistic simulation would not drastically change our conclusions.

\section{Estimate of $\Delta \phi$}
\label{sec:finest}

In this section we estimate, for the different $\tau$ decay channels, the statistical uncertainty $\Delta\phi$ 
 by taking into account  estimates 
of  measurement uncertainties and background. For the hadron-hadron 
decay channels, a judicious choice of cuts can raise the 
signal-to-background ratio to a value of $\sim 4.5$, as shown in \cite{Reinhard:2009}. 
We assume that this value can be used for all $\tau\tau$ decay 
channels and that the contribution of the background to the $\varphi_{CP}^{*}$ 
distribution is flat\footnote{We have checked for
 $e^+e^-\to Z \to \tau \tau\to \pi \pi \nu \nu$, using the formulae  for
the matrix element given e.g. in \cite{Bernreuther:1991xe},
 that this is indeed the case. This indicates that the
   $\varphi^*_{CP}$ distribution is flat, too, for the background
  reaction $e^+e^-\to Z Z$ where one $Z$ decays to $\tau\tau.$}. Furthermore, the reconstruction efficiency for signal 
events can be obtained from the analysis of \cite{Reinhard:2009} 
              (see Table~8.8 in \cite{Reinhard:2009}) to be 
$0.55$ for the combined $\tau \to \pi+\nu_{\tau}$ and $\tau \to 
\rho+\nu_{\tau}$ decay modes. We use this value to 
estimate also the   reconstruction efficiencies of the other
  decay modes. Finally, we take into account
 the  uncertainty of the impact parameter measurement by 
multiplying the asymmetry $A^{aa'}$ for all $aa'$ by $0.9$ 
(see Sect.~\ref{sec:uncert}).

\begin{table}
\begin{tabular}{|>{\centering}p{3.7cm}|>{\centering}p{1.5cm}|>{\centering}p{2.4cm}|>{\centering}p{2.3cm}|>
{\centering}p{2.3cm}|>{\centering}p{2.3cm}|}
\hline 
$\tau\tau$-decay channel & $A\,[\%]$ & \# of events
for 
${\cal L}=1\, {\rm ab}^{-1}$ & $\Delta\phi$  $[^\circ]$\\ ${\cal
  L}=1\, {\rm ab}^{-1}$ & $\Delta\phi$ $[^\circ]$\\${\cal L}=500\,
{\rm fb}^{-1}$ & $\Delta\phi$ $[^\circ]$\\${\cal L}=300\, {\rm fb}^{-1}$\tabularnewline
\hline 
$(\pi+a_{1}^{\lambda})(\pi+a_{1}^{\lambda})$ & $28.9$ & 269 & $5.5$ & $7.9$ & $10$\tabularnewline
\hline 
$\rho\rho$ & $18.0$ & $443$ & $7.0$ & $10$ & $13$\tabularnewline
\hline 
$\rho(\pi+a_{1}^{\lambda})$ & $22.8$ & $686$ & $4.4$ & $6.3$ & $8.2$\tabularnewline
\hline 
$a_{1}(\pi+a_{1}^{\lambda})$+$\rho a_{1}$+$a_{1}a_{1}$ & $10$ & 638 & $11$ & $18$ & $23$\tabularnewline
\hline 
all had-had: &  &  & $3.0$ & $4.3$ & $5.5$\tabularnewline
\hline 
\hline 
$ll$ & $4.8$ & 454 & $30$ & $36$ & $39$\tabularnewline
\hline
\hline 
$l(\pi+a_{1}^{\lambda})$ & $11.8$ & 706 & $8.7$ & $13$ & $18$\tabularnewline
\hline 
$l\rho$ & $6.0$ & $723$ & $19$ & $27$ & $31$\tabularnewline
\hline 
$la_{1}$ & $3.4$ & 292 & $38$ & $42$ & $44$\tabularnewline
\hline 
all lep-had: &  &  & 7.7 & $11$ & $15$\tabularnewline
\hline
\hline 
all: &  &  & $2.8$ & $4.0$ & $5.1$\tabularnewline
\hline 
\end{tabular}
\caption{
\label{tab:Estimate-phi}
  Estimate of the asymmetries, the number of events, and the precision 
  to the $CP$ mixing angle $\phi$ after taking into account 
  measurement uncertainties as described in the text; $\sqrt{s}=250$~GeV.
}
\end{table}

Table~\ref{tab:Estimate-phi} contains our results after 
inclusion of these effects. The asymmetries (second column) are 
reduced due to the additional flat background and due to the 
uncertainty of the impact parameter measurement. The number of 
events (for $1\, {\rm ab}^{-1}$, third column) is affected by two 
opposing effects: the limited efficiency leads to a reduction 
of signal events while  the $ZZ$ background 
increases the event numbers. The last three columns show
the resulting uncertainty $\Delta \phi$, 
for Higgs-boson production at $\sqrt{s}=250$~GeV
and for luminosities of $1\,{\rm
  ab}^{-1}$, $500\, {\rm fb}^{-1}$ 
and $300 \,{\rm fb}^{-1}$. In each case, cuts on the lepton or pion 
energies as given in Tables~\ref{tab:A_sigma_ll} to 
\ref{tab:A_sigma_rho_a1_pi_a1L} have been chosen such that the 
uncertainty  $\Delta \phi$ is minimized.

The results of Table~\ref{tab:Estimate-phi} show that the most precise
 measurement of the $CP$ mixing angle can be made using  the 
hadron-hadron decay modes. The decays $\tau^{\mp} \to \pi^{\mp},~ 
\rho^{\mp},~ a_1^{\mp,\lambda}$ have the largest impact. While 
the lepton-lepton decay modes can be neglected in the determination of
$\phi$, including the 
lepton-hadron decays in the fit can lead to an improvement of 
the precision to $\phi$ by about $8\,\%$ as compared to a fit based on 
hadron decays only. 
We finally estimate the precision of a $\phi$ measurement to be
$\Delta\phi = 2.8^{\circ}$ for a luminosity of $1\,{\rm ab}^{-1}$, 
$\Delta\phi = 4.0^{\circ}$ for $500\,{\rm fb}^{-1}$,
$\Delta\phi = 
5.1^{\circ}$ for $300 \,{\rm fb}^{-1}$, and $\Delta\phi = 
5.9^{\circ}$ for $250 \,{\rm fb}^{-1}$. 

If one considers Higgs-boson production at higher collider energies
and  assumes that  efficiencies and measurement uncertainties 
do not change significantly then one obtains the following estimates:
$\Delta\phi = 6.9^{\circ}$ for $\sqrt{s}=350$~GeV and $350 \,{\rm
  fb}^{-1}$,
$\Delta\phi = 8.8^{\circ}$ for $\sqrt{s}=500$~GeV and $500 \,{\rm fb}^{-1}$, 
and $\Delta\phi = 
14^{\circ}$ for $\sqrt{s}=1$~TeV and $1 \,{\rm ab}^{-1}$. 
The increase of   $\Delta\phi$ with increasing center-of-mass energy
is  due to the rapid decrease of the 
cross section $\sigma(e^+ e^-\to Zh)$.

\section{Conclusions}
We have studied how precisely the $CP$ nature of the 126 GeV spin-zero 
 Higgs resonance $h$, discovered last year at the LHC,  can be determined
 at a future high-luminosity linear $e^{+}e^{-}$ collider in its
 decays $h\to \tau^-\tau^+$ with subsequent decays of the $\tau$
 leptons to 1 or 3 charged prongs. The $CP$ nature of $h$ is
 reflected in the shape of the distribution of 
 an angle
 $\varphi_{CP}^{*}$  which is 
  defined in the zero-momentum frame of the charged prongs
 that result from $\tau^\pm$ decay.
 Its measurement does not require 
  the reconstruction of the $\tau$ rest frames or the
        reconstruction  of the momentum    of the   $\rho$ meson
        or  of the $a_1$ in its 1-prong decay mode.
    This  may increase the
   precision 
 with which the $CP$ mixing angle $\phi$ can be determined.

 We have analyzed all major 1- and 3-prong decays of the $\tau$
 leptons.  The most precise 
 measurement of the mixing angle $\phi$ that parameterizes the
 $CP$ nature of $h$ can be made using  the 
hadron-hadron decay modes. The decays $\tau^{\mp} \to \pi^{\mp},~ 
\rho^{\mp},~ a_1^{\mp,\lambda}$ $(\lambda=L,T)$  have the highest sensitivity to $\phi$.
 Moreover, we find that taking into account also 
  the lepton-hadron decays of $\tau^-\tau^+$ leads to an improvement of 
the precision to $\phi$ by about $8\,\%$, while the sensitivity to $\phi$
  of the modes
   $\tau^-\tau^+\to l^-l'^+$ is rather low. Assuming an integrated
   luminosity of  $1\,{\rm ab}^{-1}$ and Higgs-boson production at
   $\sqrt{s} = 250$~GeV we  estimate, including 
background and measurement uncertainties, that  the mixing angle
 $\phi$  can be determined with a statistical uncertainty of
  $\Delta\phi = 2.8^{\circ}$.
  We recall, however, that in the analysis presented here,
  several reconstruction efficiencies were assumed. Therefore,
   the achievable experimental precision on the $CP$ mixing angle 
   might be somewhat different than this number.

\section*{Acknowledgments}

We wish to thank H.~Videau for a correspondence and for providing us
with a copy of \cite{Reinhard:2009}.
The work of S.~B.\ is supported by the Initiative and Networking
Fund of the Helmholtz Association, contract HA-101 (`Physics at the
Terascale') and by the Research Center `Elementary Forces and Mathematical
Foundations' of the Johannes-Gutenberg-Universit\"at Mainz. The work
of W.~B.\ is supported by BMBF.

\section*{Appendix A}
\label{sec:Appendix-A}

The normalized distributions
of polarized $\tau$ decays to a  charged lepton $l=e,\mu$,
and to a charged pion via $\rho$ and $a_1$ decay have, in the $\tau$
rest frame,  the form 
\begin{eqnarray}
\frac{1}{\Gamma\left(\tau^{\mp} \to a^{\mp}+X\right)} 
\frac{\mbox{d}\Gamma\left(\tau^{\mp}(\hat{{\bf s}}^{\mp})
  \to a^{\mp}(q^{\mp})+X\right)}%
  {dE_{a^{\mp}} d\Omega_{a^{\mp}} / (4\pi)} 
  & = & 
  n \left(E_{a^{\mp}}\right)
  \left(1 \pm b\left(E_{a^{\mp}}\right) \, 
  \hat{{\bf s}}^{\mp} \cdot \hat{{\bf q}}^{\mp}\right) \, .
  \label{eq:dGamma_dEdOmega}
\end{eqnarray}
Here, ${\bf \hat{s}}^{\mp}$ denote the normalized spin vectors 
of the $\tau^{\mp}$ and $\hat{{\bf q}}^{\mp}$ is the direction of 
flight of $a^\mp=l^\mp, \pi^\mp$  in the respective $\tau$ rest frame. 
The spectral functions $n$ and $b$ are shown in Figs.~1a and 4 of 
Ref.\ \cite{Berge:2011ij}. The function $b(E_a)$ encodes the 
$\tau$-spin analyzing power of particle $a^\mp=l^\mp, \pi^\mp$. The 
results described in \cite{Berge:2011ij} show that for all
three decays, the function $b(E_a)$ changes sign in the allowed 
kinematic range.


\end{document}